# High sensitivity cantilevers for measuring persistent currents in normal metal rings

A. C. Bleszynski-Jayich[1], W. E. Shanks[1], B. R. Ilic[3], and J. G. E. Harris[1,2]

[1] Department of Physics, Yale University, New Haven CT, 06520, USA
[2] Department of Applied Physics, Yale University, New Haven CT, 06520, USA
[3] Cornell Nanoscale Facility, Cornell University, Ithaca NY, 14853, USA

We propose a new approach to measuring persistent currents in normal metal rings. By integrating micron-scale metal rings into sensitive micromechanical cantilevers and using the cantilevers as torque magnetometers, it should be possible to measure the rings' persistent currents with greater sensitivity than the SQUID-based and microwave resonator-based detectors used in the past. In addition, cantilever-based detectors may allow for measurements in a cleaner electromagnetic environment. We have fabricated ultra sensitive cantilevers with integrated rings and measured their mechanical properties. We present an estimate of the persistent current sensitivity of these cantilever-based detectors, focusing on the limits set by the cantilever's Brownian motion and the shot noise in the laser interferometer that monitors the cantilever.

## I. Introduction

A non-superconducting, ring-shaped conductor threaded by a magnetic flux can support a dissipationless current[1]. This current is an equilibrium property of the system, and occurs when electrons maintain their phase coherence around the ring. The prediction that measurable persistent currents should exist in realistic samples (micron-scale rings at sub-Kelvin temperatures) generated considerable interest. Later theoretical work showed that these persistent currents are also very sensitive to their electromagnetic environment[2] electron-electron interactions[3,4], and a variety of other mesoscopic phenomena[1,5,6].

However, the small magnitude of the expected current (~ 100 pA) and the need to detect the current without interrupting the ring (e.g., via its magnetic moment) have made experiments on persistent currents challenging, with only a few reported measurements [7,8,9,10,11,12,13,14]. These have been performed with SQUID-based and microwave resonator-based detectors, achieving sensitivities (referred to a persistent current) as high as 2.3 nA/Hz$^{1/2}$ [12]. These measurements have shown some agreement with theory but questions remain as to the magnitude and sign (diamagnetic or paramagnetic) of the current, the role of external electromagnetic noise, and a number of other effects.

We have developed a cantilever-based approach to measure persistent currents. Rings are patterned directly onto cantilevers and the cantilevers are used as torque magnetometers to detect the current in the rings. Sensitive cantilevers are ubiquitous in measurements of small forces; they are used in AFM[15], mass sensing[16], magnetization studies of two-dimensional electron gases[17,18], and have been used to measure the spin resonance of a single electron[19]. Cantilevers are promising for persistent currents because they offer greater sensitivity and a cleaner electromagnetic environment than achieved in previous experiments.

In this paper, we first give a short overview of persistent currents and discuss previous experimental results, their conclusions, and unresolved issues. We then describe our cantilever-based technique and estimate its sensitivity to persistent currents, focusing on limitations imposed by temperature, laser shot noise, and cantilever parameters. We describe the fabrication of high sensitivity cantilevers with integrated rings. Lastly, we compare the measured sensitivities of our cantilevers to estimates of the signals expected from persistent currents. We find promising results for the observation of persistent currents in normal metal rings.

**Overview of Persistent Currents and Previous Experimental Work**

When the phase coherence length of electrons in a normal metal ring exceeds the circumference of the ring, the single particle eigenstates extend over the whole ring and so must satisfy periodic boundary conditions. A magnetic flux $\Phi$ threading the ring modifies the boundary condition, imposing an additional phase of $\frac{2\pi\Phi}{\Phi_0}$ on the electronic wavefunctions in each traversal of the ring where $\Phi_0 = h/e$ is the flux quantum[20]. As a result the free energy $F$ is $\Phi$ dependent with periodicity $\Phi_0$ and a steady state current $I = -\frac{\partial F}{\partial \Phi}$ arises. The persistent current $I$ is also periodic in the applied flux with periodicity $\Phi_0$.

In a disorder-free ring at zero temperature, the magnitude of the persistent current is expected to $I_0 = \frac{ev_F}{L}$ where $v_F$ is the electron Fermi velocity, $e$ is the electron charge, and $L$ is the circumference of the ring. Finite temperature, scattering (both elastic and inelastic), and electron-electron interactions are all expected to modify the magnitude of the current so that

$I_p \sim \mu^*(\frac{\ell_e}{L})I_0 e^{-T/T^*}$ where $\ell_e$ is the elastic scattering length, $\mu^*$ is a parameter that accounts for the interactions, and $T^*$ is a characteristic temperature of order the Thouless energy (~ 100 mK in most experiments). Both the magnitude and the sign of $\mu^*$ (and hence the current) can vary depending on the nature of the electron-electron interactions. For a typical micron size metal ring, $I_p$ is expected to be ~ 100 pA.

The persistent current is an equilibrium phenomenon, but it can be mimicked in some regards by non-equilibrium phenomena. For example, mesoscopic rectification in a ring can convert high-frequency electromagnetic radiation to a non-equilibrium steady state DC current. This nonequilibrium current is periodic in Φ, and is diamagnetic in sign[2].

Persistent currents in ensembles of Au, Ag, Cu, and GaAs rings[8,11,12,13,14] and in single Au and GaAs rings[9,10] have been measured. The results agree with theory in some aspects, but a number of discrepancies remain unexplained. In some experiments, the magnitude of the measured currents was too large[8,911,12] and the sign inconsistent with theory[11,12]. Other questions remain as to the temperature and magnetic field dependence of the currents, the effect of spin-orbit interactions, and the role of magnetic impurities. Due to the limited number of experiments and lack of consensus on the magnitude and sign of the measured currents, further investigation seems warranted to help resolve these issues.

**Experimental Setup:**

Our measurement setup is shown in Fig. 1. Single crystal silicon cantilevers with integrated metal rings are cooled to $T = 280$ mK, the base temperature of our ³He refrigerator. A large magnetic field $\vec{B}$ is applied along the axis of the cantilever. A persistent current $I_p$ in the ring produces a magnetic moment $\vec{\mu}$ which generates a torque on the cantilever $\vec{\tau} = \vec{\mu} \times \vec{B} = \pi r^2 I_p B$ where $r$ is the radius of the ring. The cantilever's resulting deflection is monitored with an optical interferometer formed between the cantilever and the cleaved end of the optical fiber shown in the figure [21,22]. Laser light with a wavelength of 1550 nm is used. A small current-carrying coil provides magnetic flux through the ring $\Phi = \Phi_{DC} + \Phi_{AC} \sin(\omega t)$.

To estimate the magnitude of the expected signal, we start by assuming the persistent current has the form:

$$(1.1) \qquad I = I_p \sin(\frac{2\pi\Phi}{\Phi_0})$$

where $I_p$ is the magnitude of the current. The current response at $\omega$ is:

$$(1.2) \qquad I_\omega = 2I_p J_1(\frac{2\pi\Phi_{AC}}{\Phi_0})\cos(\frac{2\pi\Phi_{DC}}{\Phi_0})$$

Choosing $\Phi_{AC}$ to maximize the signal, the magnitude of $I_\omega$ is $\approx I_p$, which is expected to be ~ 0.1 nA.

The samples described in this paper are aluminum rings. Although Al is a superconductor, the rings would be in the normal state at the large $B$ used in these experiments. Persistent currents in superconductors above their critical field $H_c$ have not been accessible in the past because SQUID's and superconducting resonators are not compatible with large magnetic fields. Measurements on aluminum above $H_c$ will allow us to study persistent currents in metals with attractive electron-electron interactions and to measure how persistent currents evolve as the rings approach the superconducting transition at $H_c$.

**Sensitivity Considerations**

In the absence of technical noise (e.g., mechanical vibrations affecting the cantilever or optical/electronic noise affecting the interferometer readout), there are two important fundamental sources of noise. These are the cantilever's Brownian motion and the laser's shot noise. The cantilever's Brownian motion is driven by a white force noise with a power spectral density

$$(1.3) \quad S_{F,T} = \frac{4kk_B T_N}{\omega_o Q}$$

which sets a minimum detectable force in a bandwidth $b$ of

$$(1.4) \quad F_{min,T} = \sqrt{\frac{4kk_B T_N b}{\omega_0 Q}}$$

where $T_N$ is the cantilever's noise temperature, $Q$ its quality factor, $k$ its spring constant, and $\omega_0$ its resonant frequency. This minimum detectable force can be referred to a persistent current sensitivity via the relation

$$(1.5) \quad S_I = (\frac{1}{\pi r^2 B_\perp})^2 S_\tau = (\frac{L}{1.377})^2 (\frac{1}{\pi r^2 B})^2 S_F$$

yielding a minimum detectable current (limited by Brownian motion)

$$(1.6) \quad I_{min,T} = (\frac{L}{1.377}) \frac{1}{\pi r^2 B} \sqrt{\frac{4kk_B T_N}{\omega_0 Q}}$$

where $S_\tau$ and $S_I$ are the torque and current power spectral densities, $L$ is the length of the cantilever, $B$ is the component of the field along the axis of the cantilever, and the factor of 1.377 arises from the shape of the cantilever's lowest flexural mode [23].

Laser shot noise arises from the random arrival of photons on the photodiode monitoring the interferometer signal, and leads to a white photon flux noise with power spectral density

$$(1.7) \quad S_{SN} = \frac{2hc}{\lambda} P$$

where $c$ is the speed of light, $\lambda$ is the wavelength of the laser light, and $P$ is the laser power incident on the photodiode. Laser shot noise can be referred to a force acting on the cantilever

$$(1.8) \quad S_{F,SN} = \frac{\hbar c \lambda m^2}{4\pi P}((\omega_0^2 - \omega^2)^2 + (\frac{\omega \omega_0}{Q})^2)$$

where $m$ is the motional mass of the cantilever and $\omega$ is the measurement frequency. Again referring this to a persistent current via equation (1.5), we find this shot noise sets a limit on the minimum detectable current

$$(1.9) \quad I_{min,SN} = (\frac{L}{1.377})\frac{1}{\pi r^2 B}\sqrt{\frac{\hbar c \lambda m^2}{4\pi P}((\omega_0^2 - \omega^2)^2 + (\frac{\omega \omega_0}{Q})^2)}$$

From eqs. (1.6) and (1.9), we find the best sensitivity occurs at $\omega = \omega_0$, i.e. when the frequency of the AC component of the applied flux is tuned to the resonant frequency of the cantilever. For the laser powers used in our experiment (~ 1 nW) and the typical cantilevers we fabricate (see the next section) $I_{min,SN}$ at $\omega_0$ is 4-5 orders of magnitude less than $I_{min,T}$, indicating that the sensitivity to persistent currents will be limited by Brownian motion.

The base temperature of the refrigerator $T_b$ sets the lower bound on $T_N$. In previous work, we have shown that $T_N = T_b$ for sufficiently low laser power [our paper]. The remaining parameters in eq. (1.6) are set by the ring radius, magnetic field, and geometry and mechanical $Q$ of the cantilever. The ring radius $r$ is chosen to be small (0.3 – 0.75 µm) to maximize the persistent current (in the presence of strong disorder, the magnitude of the persistent current goes as $\frac{1}{r^2}e^{-r}$). We use $B = 9$ T, the maximum field of our magnet. For a rectangular cantilever of width $w$, thickness $t$. and length $L$, we can rewrite eq. (1.6) as

$$(1.10) \qquad I_{min,T} = \frac{1}{1.377} \frac{1}{\pi r^2 B} \left(\frac{Lwt^2}{Q}\right)^{0.5} (E\rho)^{0.25} \sqrt{k_B T_N}$$

where ρ is the density of the cantilever and $E$ is its Young's modulus. To maximize sensitivity the cantilevers should be short, narrow, and thin while retaining a high $Q$. Single crystal silicon is chosen as it has low intrinsic mechanical loss and well-established micromachining prcoesses.

We note that the presence of an AC magnetic flux will generate eddy currents $I_{eddy}$ in the ring. For a typical aluminum ring (radius 500 nm, linewidth 60 nm, resistance ~ 15 Ω) and $B_{AC}$ chosen to maximize $I_\omega$ in eq. (1.2), the magnitude of $I_{eddy}$ is ~ 0.1 pA. For the typical cantilever we fabricate (see the next section) $I_{eddy}$ is ~ 2 orders of magnitude times smaller than $I_{min,T}$ and ~ 3 orders of magnitude smaller than the expected persistent current. These eddy currents are thus expected to be negligible.

**Cantilever Fabrication:**

Figure 2 shows the cantilever fabrication steps starting from 4" diameter, <100> oriented SOI wafers with resistivity $\rho$ ~ 1,000 - 10,000 Ω-cm [ShinEtsu]. The wafers are MOS cleaned and the device layer is thinned to 330 nm through oxidation and etching in hydroflouric acid (HF). The 330 nm thickness corresponds to $3\lambda/4n_{Si}$ where $n_{Si} = 3.2$, giving maximum reflectivity. We use contact optical lithography and a $CF_4$ reactive ion etch (RIE) to define the cantilevers in the device layer (Fig. 2 (b)). The etch is timed to stop in the buried oxide (BOX) layer as $CF_4$ is not selective to Si over $SiO_2$. We use $CF_4$ because it produces relatively vertical side walls. We next pattern the aluminum rings with electron beam lithography (EBL) on a PMMA bilayer (Fig. 2 (c)). The bottom PMMA layer is ~ 300 nm thick and the top layer is ~ 50 nm thick. Alignment marks ensure registration of the rings to the cantilevers over the entire wafer. After writing the pattern and developing the PMMA we run a short oxygen plasma etch. This is followed by evaporation of 60 nm of aluminum and liftoff in a 9:1 methylene chloride/ acetone mixture. We performed overnight liftoffs followed by a few seconds of ultrasonication. The 9:1 mixture of methylene chloride and acetone was found to be much more successful than acetone alone.

The cantilevers are released in the last set of fabrication steps (Figs. 2 (d)-(f)). Windows (2 mm x 5 mm) are patterned on the backside of the wafer using contact optical lithography and the silicon handle layer is removed with a deep RIE (DRIE) using 10 μm thick photoresist as a mask. Another thick layer of photoresist protects the front side during the DRIE step (Fig. 2 (d)). After the DRIE, a ~5 min backside RIE (5 sccm $O_2$, 30 sccm $CF_4$) is performed to remove the Bosch polymer deposited during the DRIE step. This step allows the subsequent wet etchant to access the BOX layer. The BOX layer is then removed by placing the wafer, front side facing down, into 6:1 buffered oxide etch (BOE). During this step the front side remains protected with photoresist as the HF in BOE will attack aluminum. Despite the fact that BOE with surfactant (Fujistsu) was used, bubbles often form in the windows, inhibiting the etch process. These were removed by gently pipetting the solution while aiming the pipette at the bubbles.

At this point, the cantilevers are embedded in a suspended membrane of photoresist (Fig. 2 (e)). After multiple deionized water rinses, the photoresist is removed in Shipley Microposit Remover 1165 at 70° C for at least 30 minutes. The cantilevers are then transferred to isopropyl alcohol and critical point dried.

Great care is taken in keeping the cantilevers free of contamination during processing as surface dominated loss mechanisms are suspected to limit $Q$'s for cantilevers with large surface to volume ratios [24] such as the ones described in this paper.

**Results:**

A series of scanning electron micrograph (SEM) images of completed devices is shown in Fig. 3. The power spectral density of the cantilever's motion at $T_b$ = 300mK is shown in Fig. 4. The data is fit to the response function for a damped driven harmonic oscillator. The drive is a white noise force whose spectral density is given in eq. (1.3). The cantilever's $Q$, extracted from the fit, is 140,000. The area under the fit in Fig. 4 (after subtracting the baseline) is the mean square displacement of the cantilever $\langle x^2 \rangle$. By varying $T_b$ between 300 mK and 4.2 K we found that the measured values of $\langle x^2 \rangle$ have a linear dependence on $T_b$ which extrapolates to zero at $T_b = 0$ K. This confirms that the force noise driving the cantilever is thermal and $T_N = T_b$ in Fig. 4[21]. Through the equipartition theorem $\frac{1}{2}k\langle x^2 \rangle = \frac{1}{2}k_B T_N$, we determine the spring constant of the cantilever to be $k = 4 \times 10^{-4}$ N/m. The calculated force sensitivity of the cantilevers is therefore $1.6 \ aN/\sqrt{Hz}$ at 300mK. This corresponds to a current sensitivity of 20 $pA/\sqrt{Hz}$ at $B = 9$T; hence we expect to be able to measure persistent currents with cantilevers such as these.

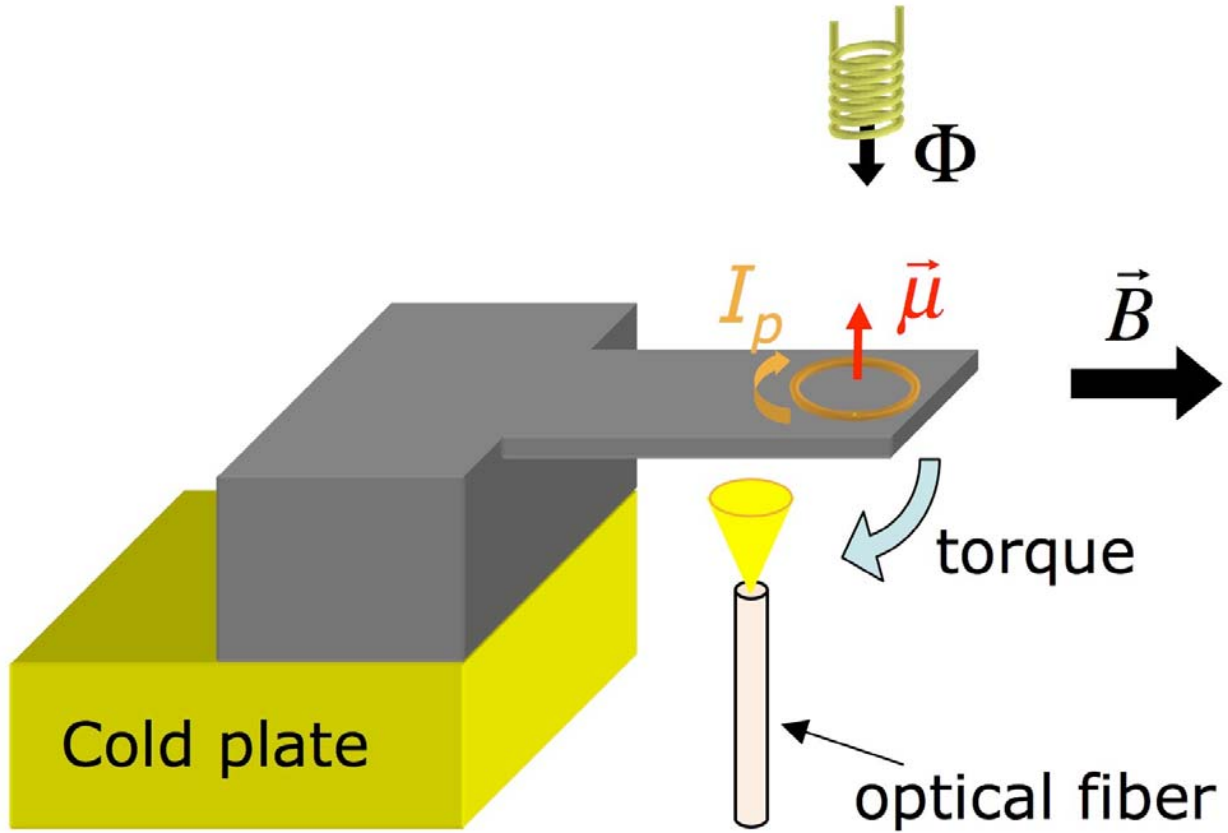

Fig 1: Cantilever torque magnetometry schematic. A metal ring is integrated onto the end of a cantilever. An applied magnetic flux Φ induces a persistent current $I_p$ in the ring, with corresponding magnetic moment $\vec{\mu}$. In the presence of a magnetic field $\vec{B}$, the cantilever experiences a torque $\vec{\tau} = \vec{\mu} \times \vec{B}$. Laser interferometry is used to monitor the resulting deflection of the cantilever. The cantilever is thermally anchored to a cold plate at $T_b \sim 300$ mK.

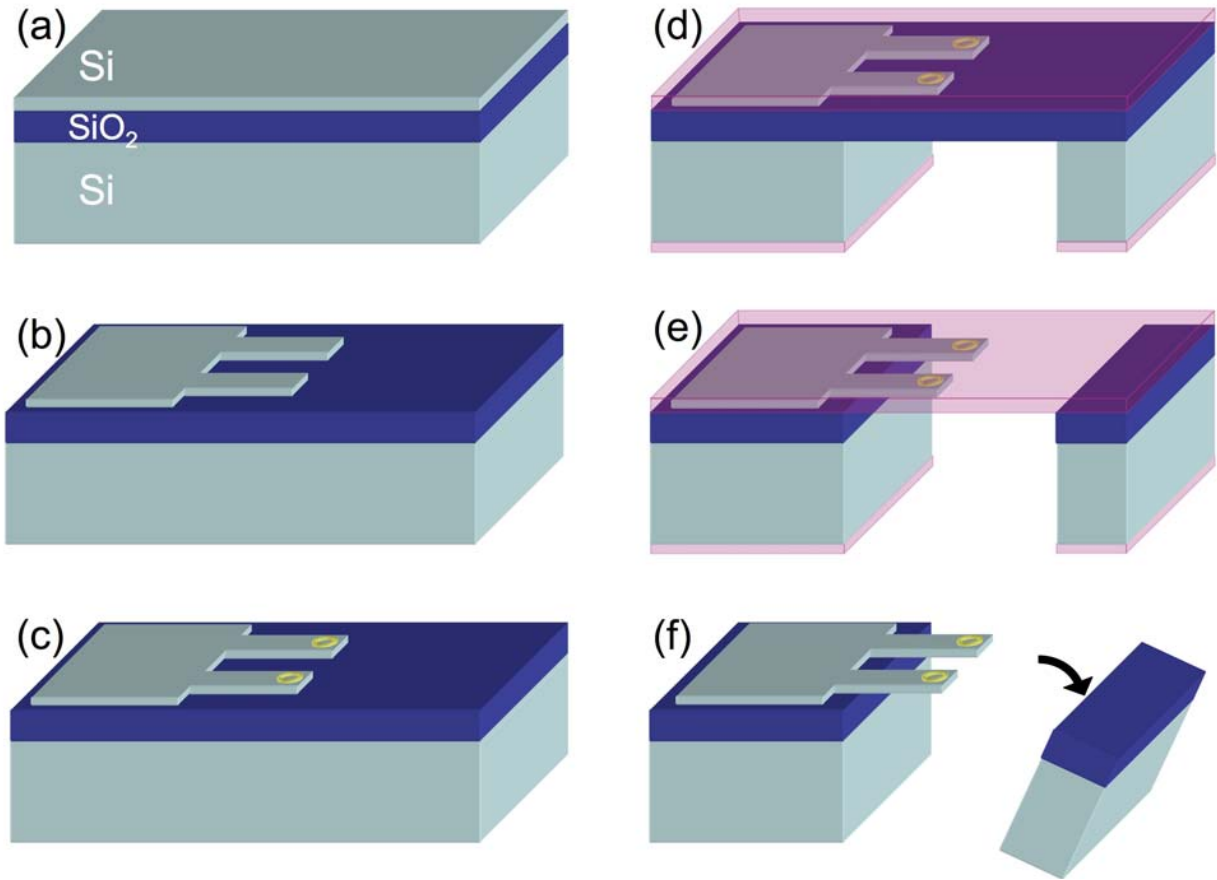

Fig 2: Fabrication of cantilevers with integrated metal rings. (a) The process starts with an SOI wafer. (b) Photolithography and a $CF_4$ etch define the cantilevers in the device layer. (c) Rings are patterned onto the end of the cantilevers with electron beam lithography, aluminum evaporation, and liftoff. (d) Windows on the backside of the wafer are formed with photolithography and a deep reactive ion etch. Photoresist protects the front side during this step. (e) The BOX layer in the exposed window is removed with 6:1 BOE. (f) The frontside photoresist is removed and the chip supporting the cantilevers is cleaved from the rest of the wafer.

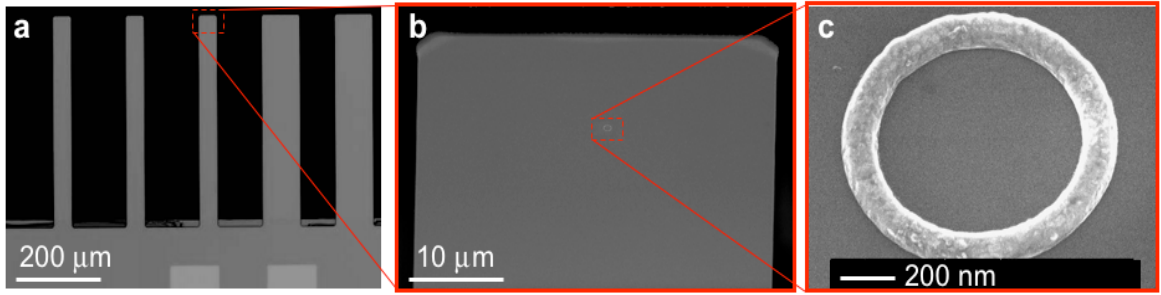

Fig 3: SEM micrographs of completed cantilevers with integrated Al rings. (a) Five cantilevers of varying width (40 μm and 80 μm) and length 425 μm. (b) A magnified view of the center cantilever in (a) showing the metal ring near the end of the cantilever. (c) A further magnified view of the cantilever in (b) showing an evaporated aluminum ring.

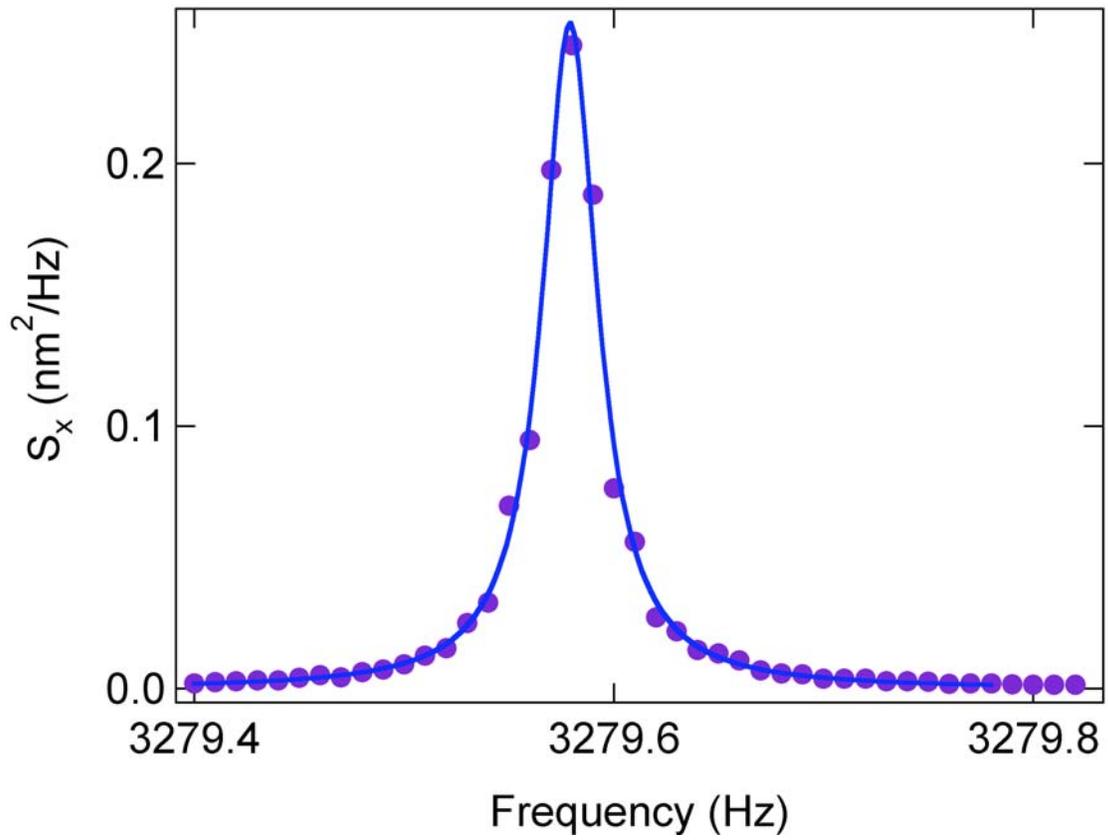

Fig 4: Displacement power spectral density of an undriven cantilever's Brownian motion at $T_b$ = 300 mK. The Lorentzian fit yields a cantilever $Q$ = 140,000 and a spring constant of 4 x $10^{-4}$ N/m.